\documentclass[12pt]{article}
\usepackage{authblk}

\usepackage{xcolor, tcolorbox}  % Color text
\newcommand{\Red}{\color{red}}

\usepackage[normalem]{ulem} % striketrhough text, use \sout{text}
\usepackage[labelfont=bf]{caption}

\usepackage[utf8]{inputenc}
\usepackage[margin=1in]{geometry}

\usepackage{amsmath, amssymb}

\usepackage{graphicx}       % For including graphics
\usepackage{subcaption}     % For subfigures
\usepackage{caption}        % For custom caption formatting
\usepackage{cite}

\DeclareCaptionFormat{myformat}{#1#2#3\par\rule{\linewidth}{0.4pt}}
\usepackage[percent]{overpic}

\usepackage{tcolorbox}

\usepackage{fancyhdr}
\usepackage[super,sort&compress]{natbib}
\setcitestyle{comma}

\usepackage{hyperref}

\usepackage{romannum}

\usepackage[nopatch]{microtype}

\usepackage{booktabs}

\title{\textbf
{CO$_2$-Induced Shifts in the Confined Phase Transformation of CH$_4$ in Metal-Organic Frameworks}
}
\author[1,2,*]{Anthony Dorhauer}
\author[3]{Bartosz Mazur} 
\author[3,4]{Bogdan Kuchta}
\author[1,2,*]{Carlos Wexler}

\affil[1]{\small Department of Physics and Astronomy, University of Missouri, USA}
\affil[2]{\small Materials Science and Engineering Institute, University of Missouri, USA}
\affil[3]{\small Department of Micro, Nano, and Bioprocess Engineering, Wrocław University of Science and Technology, Poland}
\affil[4]{\small  MADIREL, CNRS, Aix-Marseille University, France}
\affil[*]{\small  Corresponding authors: ajdfbb@missouri.edu, wexlerc@missouri.edu}

\date{August 2026}

\begin{document}

\pagenumbering{arabic}
\maketitle
%%%%%%%%%%%%%%%%%%%%%%%%%%%%%%%%%%%%%%%%%%%%%%%%%%%%%%%%%%%%%%%%%%%%%%%%%%%%%%%
\begin{abstract}
\noindent
A sharp increase in gas uptake over a narrow pressure range can occur for fluids confined in nanoporous materials such as metal–organic frameworks (MOFs), reflecting a transformation between distinct adsorption states, often accompanied by the existence of metastable states. Understanding how co-adsorbed species modify this behavior is important for predicting and optimizing gas storage and separation. Here, we investigate the effect of CO$_2$ on the adsorption transformation of confined CH$_4$ in rigid IRMOF-1 and IRMOF-8. Transition-Matrix Monte Carlo (TMMC) simulations were performed for systems containing 0-75 and 0-150 CO$_2$ molecules in IRMOF-1 and IRMOF-8, respectively. While the addition of CO$_2$ decreases the maximum CH$_4$ uptake, consistent with competition for pore space, remarkably, the presence of CO$_2$ significantly shifts the discontinuity in the CH$_4$ adsorption isotherm to substantially lower pressures relative to the pure CH$_4$ system. This promotion of the high-density CH$_4$ phase at lower pressures demonstrates that CO$_2$ substantially alters the thermodynamic landscape governing confined CH$_4$ through changes in both fluid-fluid and fluid-framework interactions. These results establish CO$_2$ as a thermodynamic modifier of confined CH$_4$ behavior and suggest that controlled co-adsorption provides a means of tuning adsorption transformations in nanoporous materials.
\end{abstract}

\begin{center}
{\bf Supplemental Information and Input Files available upon request.}
\end{center}

%%%%%%%%%%%%%%%%%%%%%%%%%%%%%%%%%%%%%%%%%%%%%%%%%%%%%%%%%%%%%%%%%%%%%%%%%%%%%%%
\section{Introduction}

Metal–organic frameworks (MOFs) provide well-defined model systems for studying the fundamental physics of fluids under nanoconfinement. Their exceptionally high surface areas and tunable pore geometries create controlled environments in which fluid-fluid and fluid-framework interactions can produce behavior that differs markedly from bulk thermodynamics \cite{Kuchta2017, Mason2015, Neimark2006, Morishige2006, Dorhauer2025}. MOFs readily host guest molecules such as methane (CH$_4$) and carbon dioxide (CO$_2$), making them particularly useful for investigating how confinement, composition, and interaction strength influence adsorption, phase behavior, and collective structural transformations \cite{Morishige2006, Ortiz2014, Li2025, Kuchta2017,Dorhauer2025,Stankiewicz2026}.

Recent numerical studies of CH$_4$ adsorption in IRMOF-1 \cite{Mazur2022} and IRMOF-8 \cite{Dorhauer2025} under subcritical conditions, along with experimental investigations of CO$_2$ adsorption in IRMOF-1 \cite{Walton2008}, have revealed a structural transformation of the adsorbed phase driven by the increasing importance of fluid–fluid interactions relative to fluid–framework interactions. This transformation manifests as a pronounced discontinuity in the adsorption isotherm, indicative of a cooperative reorganization of the confined fluid \cite{Kuchta2017, Mazur2022, Dorhauer2025, Stankiewicz2026}. 
Transition-Matrix Monte Carlo (TMMC) simulations provide direct access to the underlying free energy landscape \cite{Fitzgerald1999, Errington2003, Shen2004, Siderius2013}, revealing metastable states associated with distinct structural arrangements of the adsorbate \cite{Dorhauer2025}. Despite extensive investigations of single-component adsorption in MOFs, comparatively little is known about how co-adsorbed species alter the free energy landscape governing the phase transformations in these confined spaces. In particular, the influence of CO$_2$ on the phase behavior of subcritical CH$_4$ has not been systematically examined.

CH$_4$ constitutes approximately 85–90\% of commercial natural gas, while CO$_2$ is commonly found in certain sources \cite{BritannicaNaturalGas}. Thus, the effect of CO$_2$ impurities on CH$_4$ adsorption is directly relevant to natural gas storage and separation technologies \cite{Mason2014}. Beyond these practical considerations, controlled addition of CO$_2$ provides a means of perturbing the thermodynamic landscape of a confined fluid and probing how multicomponent interactions influence adsorption-induced transformations in MOFs \cite{Li2009}. Understanding this coupling is therefore important both for optimizing MOF-based gas storage and separation and for developing a fundamental understanding of multicomponent confined-fluid behavior \cite{Mason2014, Li2009, Sumida2012}.

In this work, we investigate the co-adsorption of CH$_4$ and CO$_2$ in rigid IRMOF-1 and IRMOF-8 using Transition-Matrix Monte Carlo (TMMC) simulations. These frameworks provide a useful comparison because previous studies have shown that they exhibit qualitatively different free energy landscapes, with IRMOF-8 possessing an additional intermediate metastable state that is absent in IRMOF-1 \cite{Dorhauer2025}. Systems containing varying CO$_2$ loadings, ranging from 0–75 molecules in IRMOF-1 and 0–150 molecules in IRMOF-8, were studied over the temperature range 80–130 K. We examine specifically how CO$_2$ influences CH$_4$ adsorption, metastability, and the free energy landscape associated with the confinement-induced transformation of CH$_4$. By comparing systems containing various amounts of CO$_2$ with simulations employing modified interaction parameters, we further identify the physical mechanisms through which co-adsorbed CO$_2$ alters the thermodynamics and phase behavior of confined CH$_4$.

This work demonstrates that CO$_2$ simultaneously suppresses the maximum CH$_4$ uptake while promoting the transformation from a low-density to a high-density phase at lower pressure, establishing co-adsorption as a means of tuning the thermodynamics and phase behavior of confined fluids in nanoporous materials.

% MOVE CLOSER TOGETHER (CROP PDFs)
\begin{figure}[t]
    \centering
\includegraphics[width=0.8\textwidth]{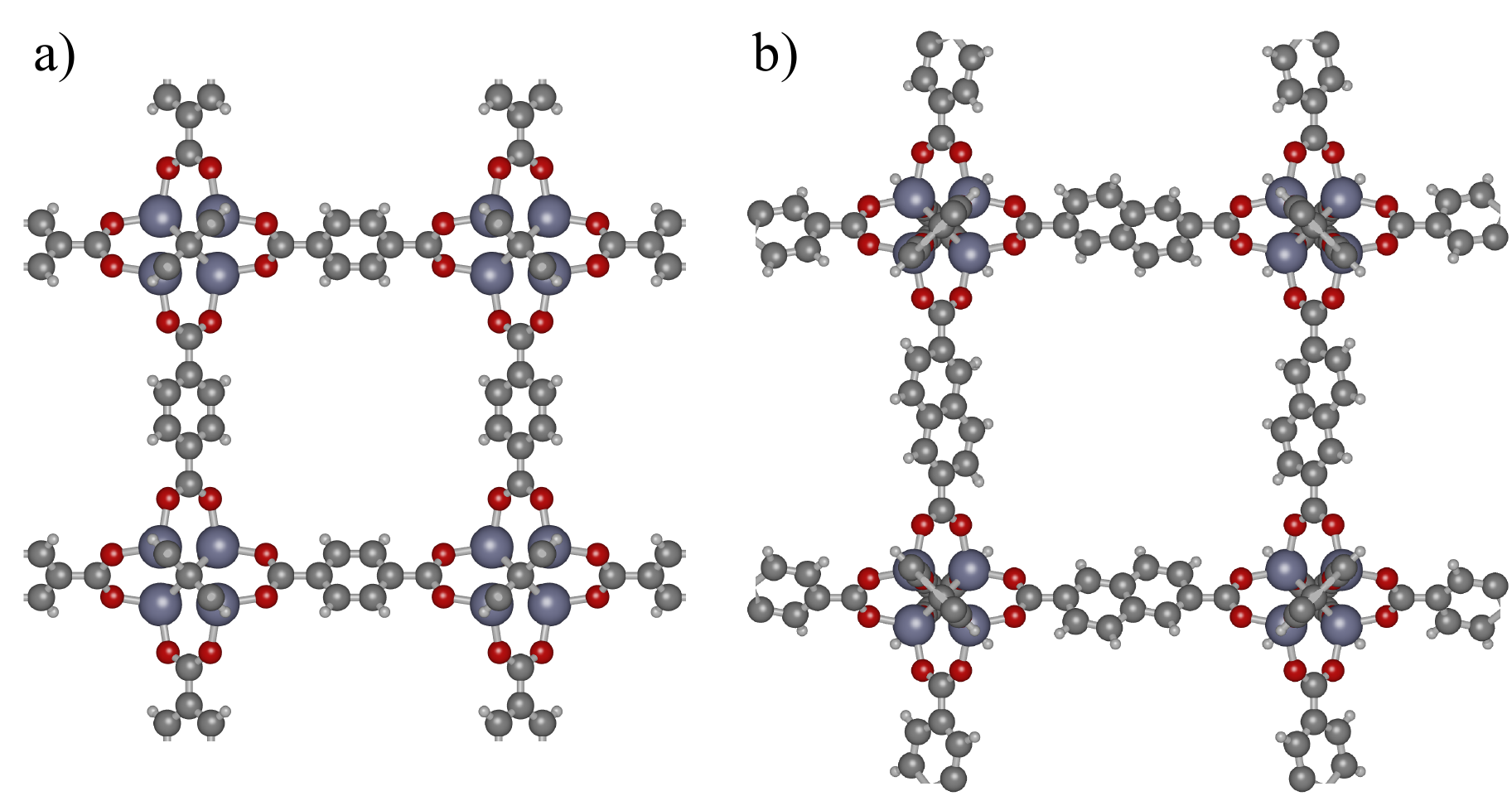}
    % % First subfigure
    % \begin{subfigure}[b]{0.45\textwidth}
    %     \begin{overpic}[width=0.87\textwidth]{Plots/PDFs/IRMOF-1.pdf}
    %         \put(2,92){\textbf{a)}}
    %     \end{overpic}
    %     \label{fig:IRMOF-1}
    % \end{subfigure}
    % \hfill
    % % Second subfigure
    % \begin{subfigure}[b]{0.45\textwidth}
    %     \begin{overpic}[width=\textwidth]{Plots/PDFs/IRMOF-8.pdf}
    %         \put(2,92){\textbf{b)}}
    %     \end{overpic}
    %     \label{fig:IRMOF-8}
    % \end{subfigure}
    \caption{(a) IRMOF-1 unit cell with dimensions 25.832\AA $\, \times$ 25.832 \AA $\, \times$ 25.832\AA. (b) IRMOF-8 unit cell with dimensions 30.0195 \AA $\, \times$ 30.0915 \AA $\, \times$ 30.0915 \AA. Atom colors: Zn (purple), O (red), C (dark gray), H (light gray).}
    \label{fig:IRMOFs-1,8}
\end{figure}

%%%%%%%%%%%%%%%%%%%%%%%%%%%%%%%%%%%%%%%%%%%%%%%%%%%%%%%%%%%%%%%%%%%%%%%%%%%%%%%
\section{Computational Methods}

Our simulations were performed using single unit cells of IRMOF-1 and IRMOF-8 (Fig. \ref{fig:IRMOFs-1,8}). The frameworks were held rigid throughout all simulations; this approximation substantially reduces computational cost while preserving the adsorption-induced structural transformations previously reported in both systems \cite{Mazur2022, Dorhauer2025}. Previous studies incorporating framework flexibility have shown that flexibility can modify energy barrier heights and transition pressures; however, it does not eliminate the cooperative adsorption behavior associated with large, interconnected pores \cite{Mason2014, Coudert2015}. The framework structural models were obtained from experimental crystallographic data \cite{Eddaoudi2002, Feldblyum2012, Feldblyum2012b}. IRMOF-1 and IRMOF-8 belong to the isoreticular metal-organic framework (IRMOF) series and therefore share a common Zn$_4$O secondary building unit. The two frameworks differ primarily in their organic linkers: IRMOF-1 employs 1,4-benzenedicarboxylate linkers, whereas IRMOF-8 utilizes naphthalene-2,6-dicarboxylate linkers, resulting in a larger pore volume and larger unit-cell dimensions \cite{Yaghi2003}.

Interactions between molecules were modeled using a combination of electrostatic (Coulomb) and 12-6 Lennard-Jones (LJ) potentials

\begin{equation} \label{eq:U_Coulomb}
    U_{C}(r) = \frac{q_i q_j}{4 \pi \epsilon}\frac{1}{r_{ij}}
\end{equation}

\begin{equation} \label{eq:U_LJ}
    U_{LJ}(r) = 4 \varepsilon_{ij} [(\frac{\sigma_{ij}}{r_{ij}})^{12} - (\frac{\sigma_{ij}}{r_{ij}})^{6}]
\end{equation}
where $\varepsilon_{ij}$ is the depth of the energy well, $\sigma_{ij}$ is the distance at which $U_{LJ}(r)$ is zero, $r_{ij}$ is the distance between atoms $i$ and $j$, $q$ is the atomic charge, and $\epsilon$ is the dielectric constant. Cross-interactions were calculated using Lorentz-Berthelot mixing rules: $\varepsilon_{ij} = \sqrt{\varepsilon_i \varepsilon_j}$; $\sigma_{ij} = (\sigma_i + \sigma_j)/2$ \cite{Lorentz1881, Berthelot1898}. Ewald summation was used to calculate long-range electrostatic interactions \cite{Ewald1921}.

Following previous studies of IRMOF-1 and IRMOF-8, Zn atoms were described using the Universal Force Field (UFF) parameters \cite{Rappe1992}, whereas framework C, O, and H atoms were modeled using DREIDING parameters \cite{Mayo1990}. CH$_4$ was modeled as a single-site pseudo-atom in the United Atom Approximation (UAA) \cite{Martin1998} with a molecular mass of 16.04246 amu, parametrized by the diffuse-accessible-center nonbonded interaction sites (DACNIS) model \cite{Martin2001}. 
CO$2$ was modeled using the TraPPE force field \cite{Potoff2001}, with Coulombic and Lennard-Jones interactions assigned to the C and O sites and a fixed C–O bond length of $d_{CO}=1.149$ \AA, resulting in an electric quadrupole moment $Q_{zz}^{CO_2} = -14.7 \times 10^{-40}$ $\mathrm{C\,m^2}$, comparable to experimental values of $-14 \times 10^{-40}$ to $-15 \times 10^{-40}$ $\mathrm{C\,m^2}$. \cite{Yachmenev2021}
The C–O bonds were constrained to remain rigid. Atom specific parameters are listed in \textbf{Table \ref{table:Coulomb}} and \textbf{Table \ref{table:LJ}}.

\begin{table}[h!]
\caption{Electrostatic parameters used in simulations. For more details see \textbf{SI-6} in the Supporting Information {\Red [cite to URL]}.}
\label{table:Coulomb}
\centering
\begin{tabular}{|c|c|c|}
\hline

% ---- Molecules section ----
\textbf{Atom} & \multicolumn{2}{c|}{\textbf{Atom Charge ($e$)}} \\ \hline

CH$_4$ & \multicolumn{2}{c|}{0.00} \\ \hline
C$_{\mathrm{CO_2}}$ & \multicolumn{2}{c|}{0.70} \\ \hline
O$_{\mathrm{CO_2}}$ & \multicolumn{2}{c|}{-0.35} \\ \hline

\textbf{} & \textbf{IRMOF-1} & \textbf{IRMOF-8} \\ 
\hline
Zn & 0.973 & 0.978 \\ \hline
O & -1.036/-0.541 & -1.047/-0.564 \\ \hline
C & 0.591/-0.018/-0.111 & 0.616/-0.034/-0.138 \\ \hline
H & 0.128 & N/A \\ \hline
\end{tabular}
\end{table}

\begin{table}[h!]
\caption{Lennard-Jones parameters used in simulations. 
\label{table:LJ}}
\centering
\begin{tabular}{|c|c|c|}
\hline
\textbf{Atom} & $\varepsilon$ (K) & $\sigma$ (\AA) \\ \hline
CH$_4$ & 158.50 & 3.65 \\ \hline
C$_{\mathrm{CO_2}}$ & 27.00 & 2.80 \\ \hline
O$_{\mathrm{CO_2}}$ & 79.00 & 3.05 \\ \hline
Zn & 62.40 & 2.46 \\ \hline
O & 48.16 & 3.03 \\ \hline
C & 47.86 & 3.47 \\ \hline
H & 7.65 & 2.85 \\ \hline
\end{tabular}
\end{table}

All simulations were run using a modified version of the RASPA2 software \cite{Mazur2025, Mazur2024, Dubbeldam2016}. We applied periodic boundary conditions in all three directions with a spherical cutoff radius of $r_c = 12.8$~\AA\ in IRMOF-1 and $r_c =14$~\AA\ in IRMOF-8 for the LJ interactions, with both cutoff radii smaller than half the corresponding unit-cell length. Fugacities for both CH$_4$ and CO$_2$ were obtained using the Peng-Robinson equation of state \cite{Peng1976}. The critical temperatures, critical pressures, and acentric factors used for CH$_4$ were $T_c = 190.564$ K, $P_c = 4599.2$ kPa, and $\omega = 0.01142$, respectively, while those for CO$_2$ were $T_c = 304.128$ K, $P_c = 7377.3$ kPa, and $\omega = 0.22394$. TMMC simulations were run for 600,000 cycles with 60,000 initialization cycles in IRMOF-1, and 1,000,000 cycles along with 100,000 initialization cycles in IRMOF-8 due to the added complexity of the free energy landscape. For each simulation, the number of CO$_2$ molecules was fixed, while CH$_4$ molecules were exchanged with a reservoir at fixed chemical potential. Thus, CO$_2$ was sampled canonically, while CH$_4$ was sampled grand canonically through the TMMC procedure.

% --- SHORTEN TMMC EXPLANATION USING THIS --- %
TMMC simulations were used to obtain the macrostate probability distribution, $\Pi(N)$, and the corresponding free energy profile, $\Omega(N)$. Following the standard formulation of Fitzgerald {\em et al.} and Errington \cite{Fitzgerald1999, Errington2003, Shen2004, Siderius2013}, transition probabilities between neighboring macrostates were accumulated using ghost insertion and deletion moves, while all configurational sampling was performed in the canonical ensemble. 

% We employed the TMMC method \cite{Fitzgerald1999, Errington2003, Shen2004, Siderius2013} to obtain the macrostate-dependent free energy profiles $\Omega(N)$. The TMMC method runs simulations in the \textit{canonical} ensemble (fixed $N$), and applies the "ghost particle insertion/deletion move" in which particles are artificially inserted/deleted ($N \rightarrow{N \pm{1}}$); acceptance probabilities are calculated and updated, then the moves are always rejected. The acceptance probabilities are given by:

\begin{equation} \label{eq:p_insertion}
    p_{acc}(N \rightarrow{N+1}) = \min\left(1,\frac{V \Lambda^{-3}}{N+1} e^{-\beta (U_{N+1} - U_N - \mu)}\right),
\end{equation}

\begin{equation} \label{eq:p_deletion}
    p_{acc}(N \rightarrow{N-1}) = \min\left(1,\frac{N \Lambda^3}{V} e^{-\beta (U_N - U_{N-1} + \mu)}\right),
\end{equation}
Where $\Lambda = $ is the thermal de Broglie wavelength and the chemical potential, $\mu$, is determined by the pressure (or fugacity $f$; our fugacity coefficient was 1), by the relation $\mu(T,f) = \mu^0(T) + k_BT \ln(f/f^0)$, and $\beta = 1 / k_B T$. These acceptance probabilities were accumulated in a collection matrix and subsequently normalized to obtain the transition probabilities $P(N \rightarrow N\pm1)$ between neighboring macrostates.

\begin{equation} \label{eq:free_energy}
    \Delta \Omega(N \rightarrow{N + 1}) \equiv \Omega(N + 1) - \Omega(N) = -k_B T \ln{\frac{P(N \rightarrow{N + 1})}{P(N + 1 \rightarrow{N})}}.
\end{equation}

The macrostate free energy, expressed as the grand potential $\Omega(N)$, was obtained by integrating the free-energy differences between neighboring macrostates. Histogram reweighting was used to obtain free-energy profiles at arbitrary chemical potentials without additional simulations:

\begin{equation} \label{eq:reweighting}
    \ln \Pi(N; \mu, V, T) = \ln \Pi(N; \mu_0, V, T) + \beta N(\mu - \mu_0) + C,
\end{equation}
where, $\mu_0$ is the known chemical potential, $\mu$ is the desired chemical potential, $\beta N(\mu - \mu_0)$ is the reweighting term, and $C$ is a constant independent of $N$. Adsorption isotherms were obtained by locating the global minimum of each free-energy profile (\textbf{SI-1}), corresponding to the equilibrium CH$_4$ uptake at a given pressure (or fugacity).

%%%%%%%%%%%%%%%%%%%%%%%%%%%%%%%%%%%%%%%%%%%%%%%%%%%%%%%%%%%%%%%%%%%%%%%
\section{Results and Discussion}

In general, the isotherms shown in \textbf{Figure \ref{fig:Isotherms_IRMOFs-1,8}} exhibit an initial, gradual increase in CH$_4$ uptake, followed by a sharp rise over a narrow pressure range. This abrupt transition is governed by the free energy barriers separating distinct minima in the underlying free energy landscape and is reminiscent of a vapor-liquid phase transition under confinement. IRMOF-1 (\textbf{Figure \ref{fig:Isotherms_IRMOFs-1,8}}, top row) displays bimodal behavior, corresponding to low- and high-density states \cite{Mazur2022}. In contrast, IRMOF-8 (\textbf{Figure \ref{fig:Isotherms_IRMOFs-1,8}}, bottom row) reveals the emergence of a third, intermediate metastable state, indicative of a more complex free energy landscape \cite{Dorhauer2025}.

\begin{figure}[t]
    \centering
    % Row 1
    \begin{subfigure}[b]{0.32\textwidth}
        \begin{overpic}[width=\textwidth]{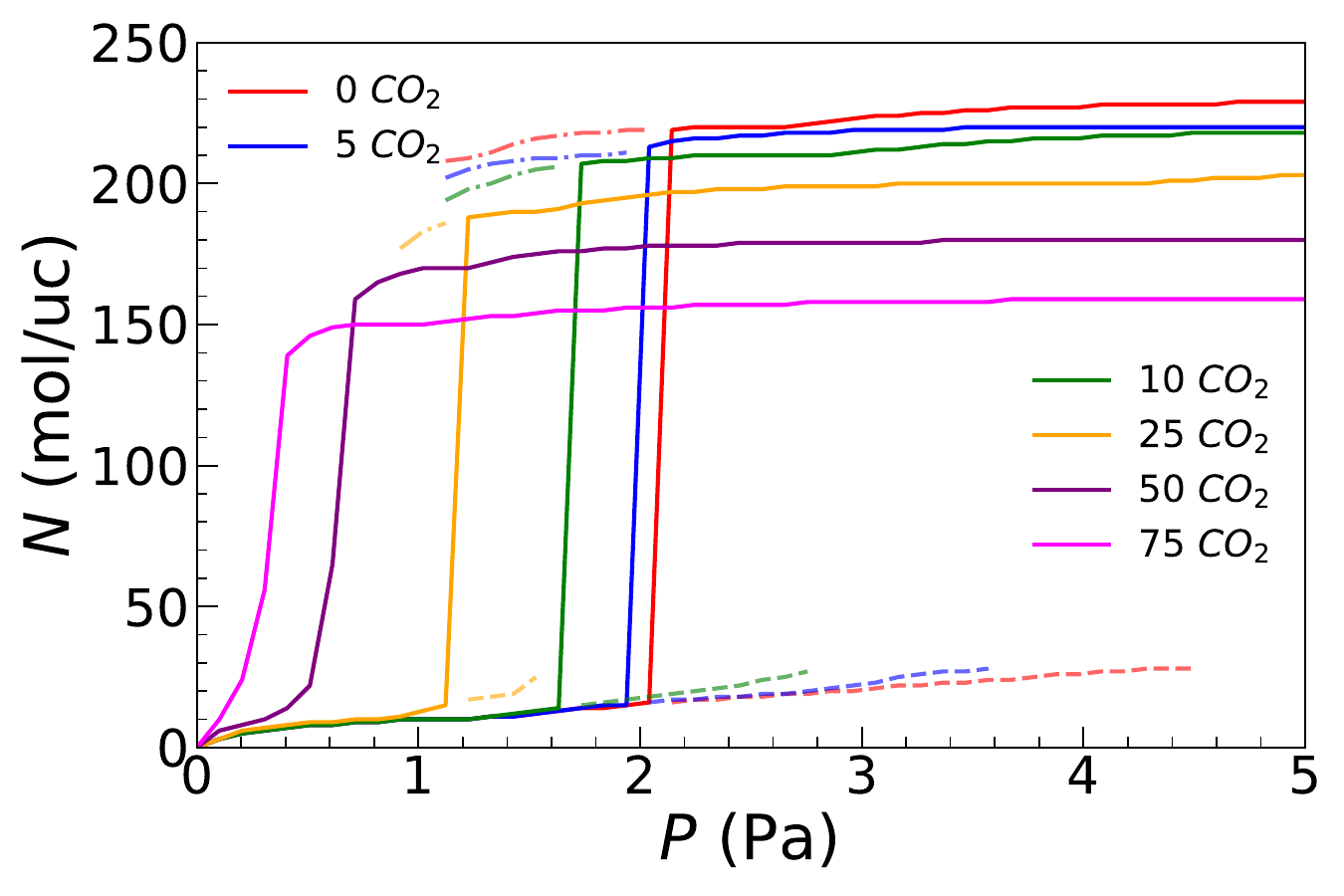}
            \put(0,70){\textbf{a)}}
        \end{overpic}
        \label{fig:Isotherms_IRMOF-1_80K}
    \end{subfigure}
    \hfill
    \begin{subfigure}[b]{0.32\textwidth}
        \begin{overpic}[width=\textwidth]{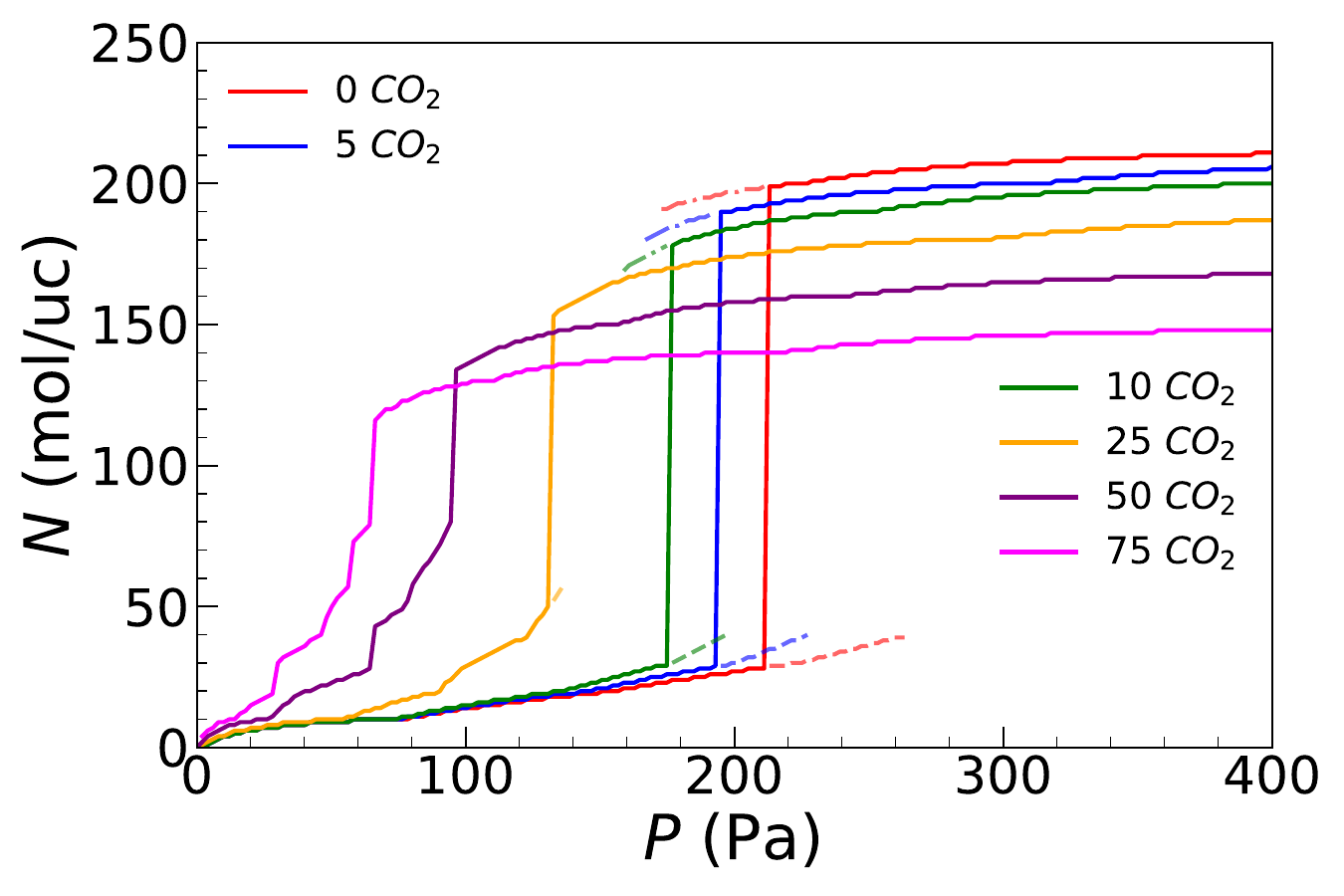}
            \put(0,70){\textbf{b)}}
        \end{overpic}
        \label{fig:Isotherms_IRMOF-1_102K}
    \end{subfigure}
    \hfill
    \begin{subfigure}[b]{0.32\textwidth}
        \begin{overpic}[width=\textwidth]{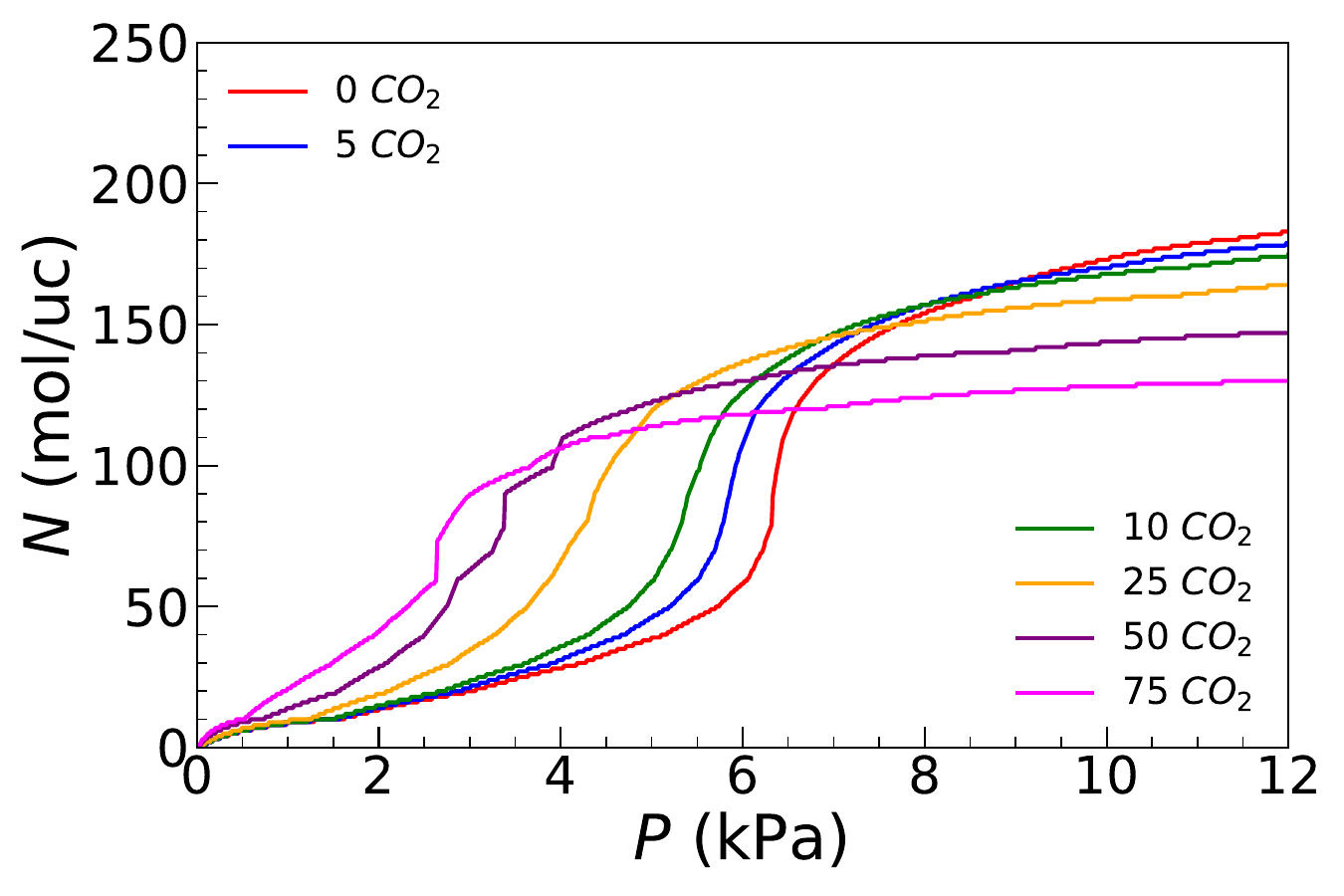}
            \put(0,70){\textbf{c)}}
        \end{overpic}
        \label{fig:Isotherms_IRMOF-1_130K}
    \end{subfigure}
    \vspace{0.5cm}
    % Row 2
    \begin{subfigure}[b]{0.32\textwidth}
        \begin{overpic}[width=\textwidth]{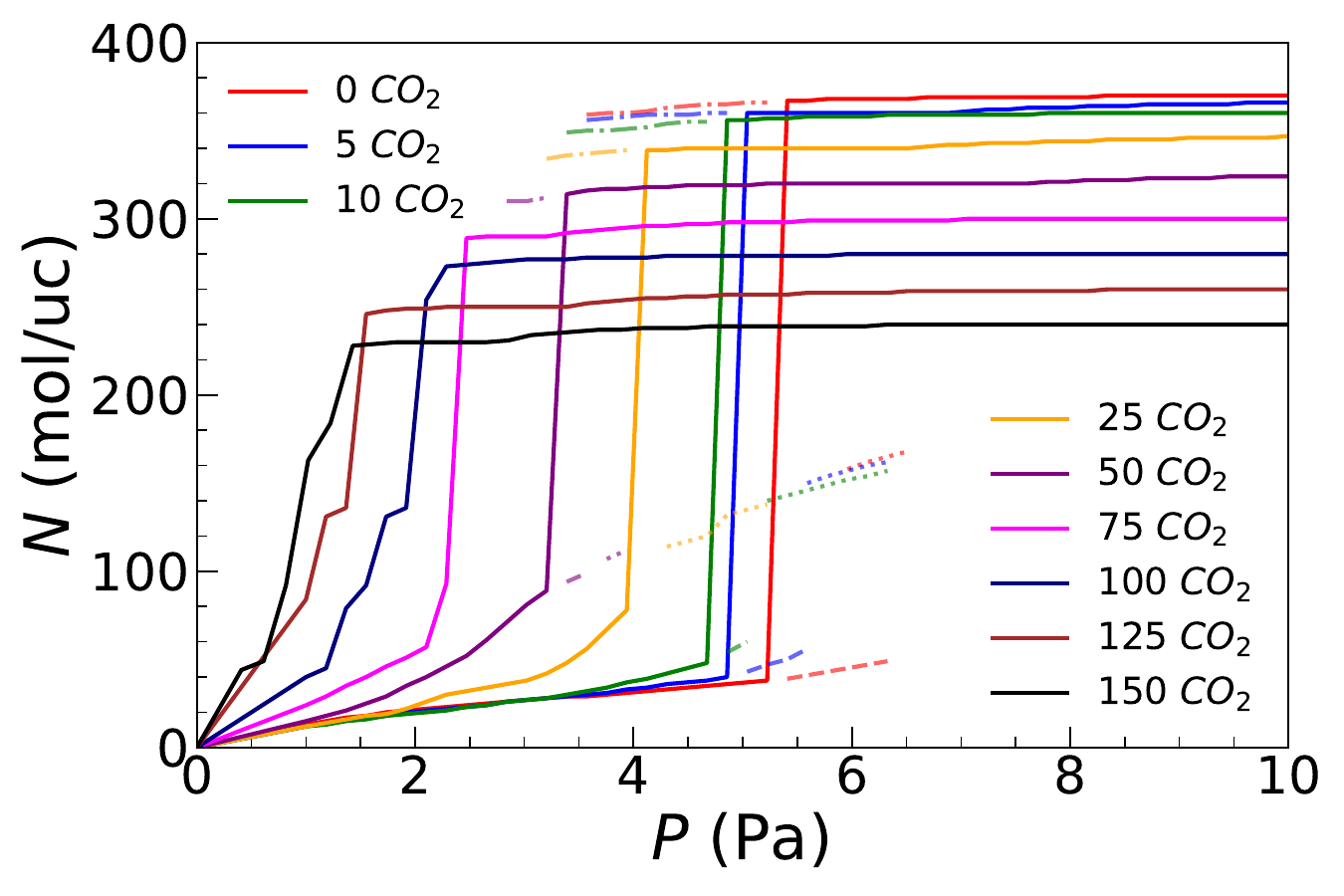}
            \put(0,70){\textbf{d)}}
        \end{overpic}
        \label{fig:Isotherms_IRMOF-8_80K}
    \end{subfigure}
    \hfill
    \begin{subfigure}[b]{0.32\textwidth}
        \begin{overpic}[width=\textwidth]{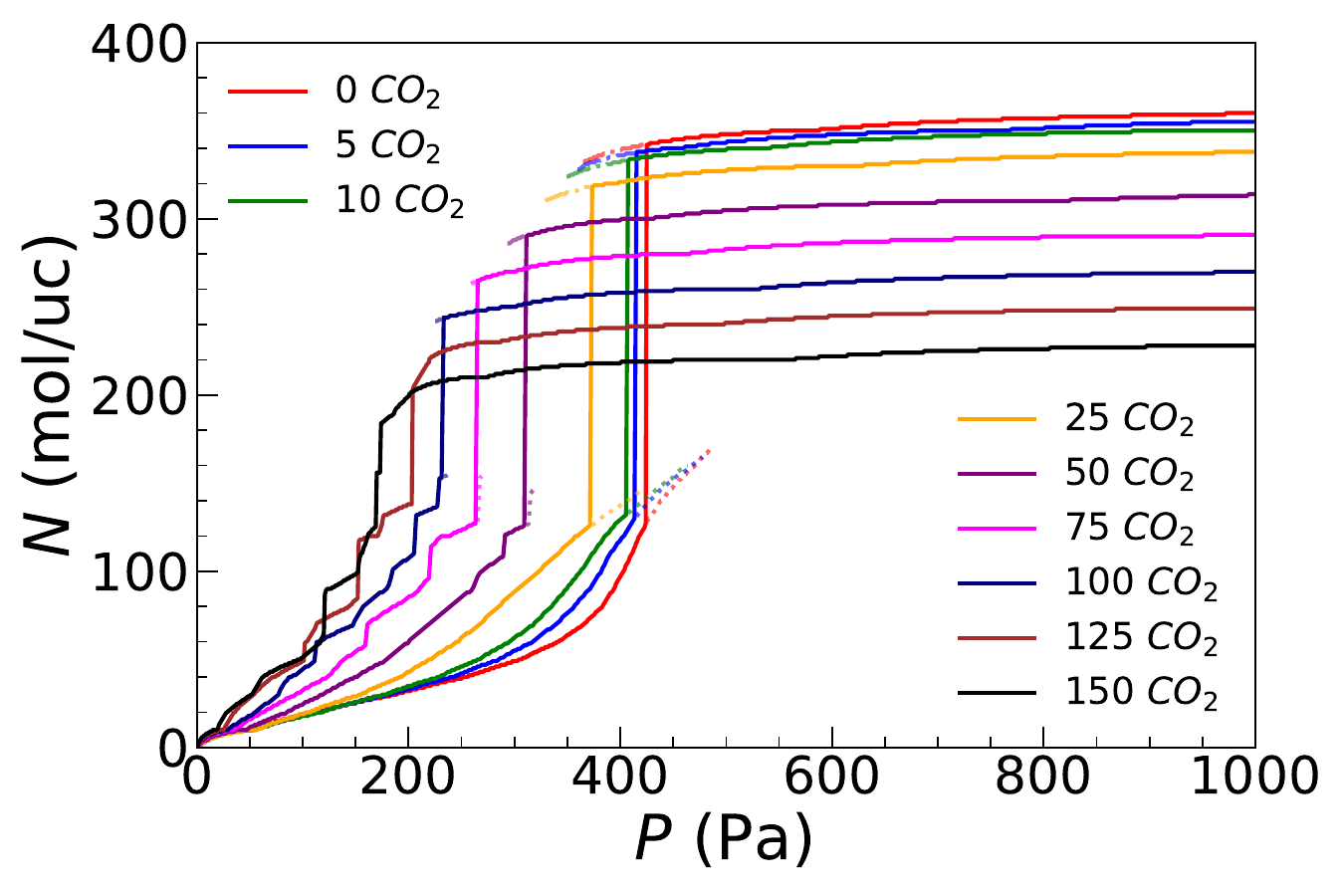}
            \put(0,70){\textbf{e)}}
        \end{overpic}
        \label{fig:Isotherms_IRMOF-8_102K}
    \end{subfigure}
    \hfill
    \begin{subfigure}[b]{0.32\textwidth}
        \begin{overpic}[width=\textwidth]{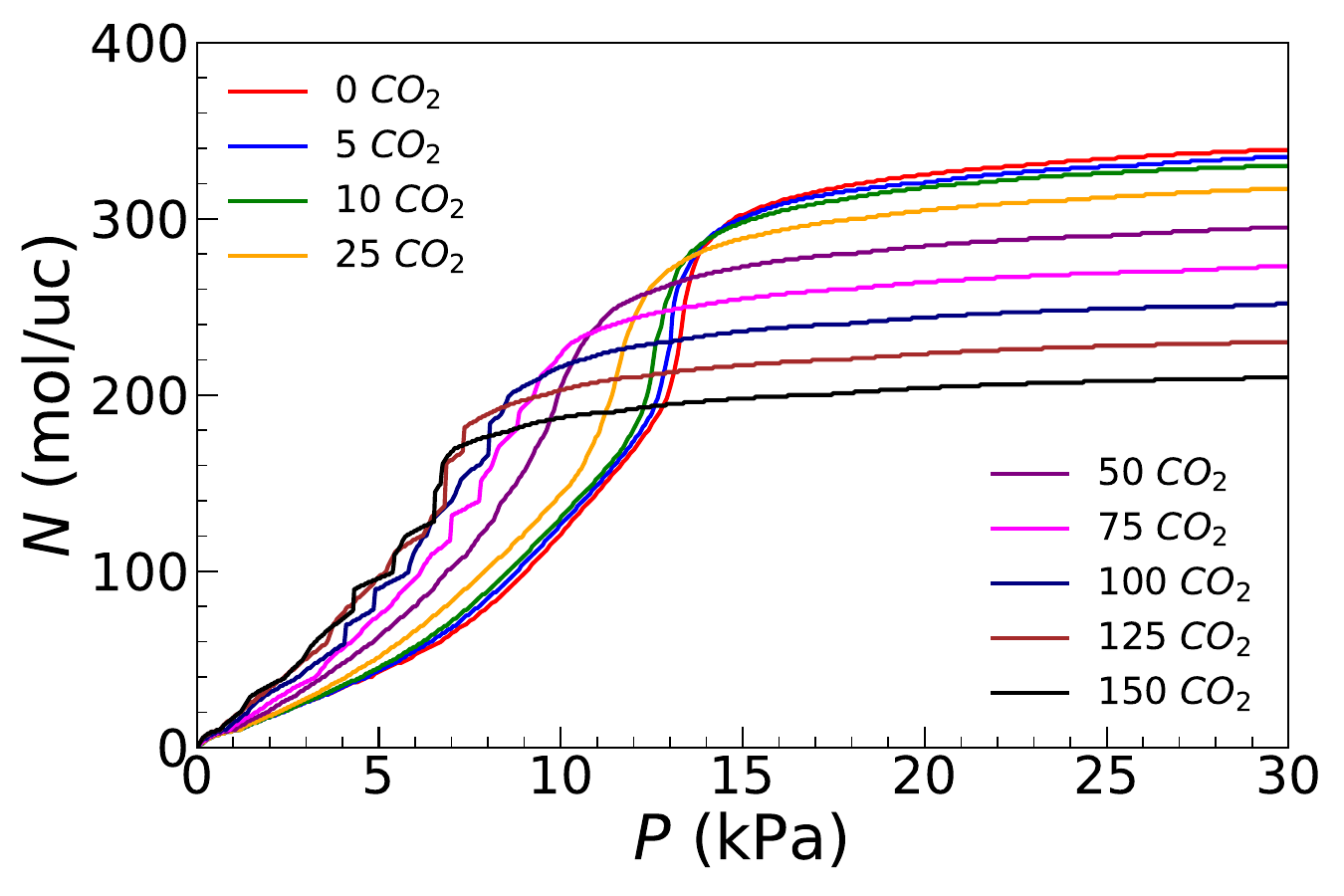}
            \put(0,70){\textbf{f)}}
        \end{overpic}
        \label{fig:Isotherms_IRMOF-8_130K}
    \end{subfigure}
    \caption{Adsorption isotherms for CH$_4$ in: 
    (a) IRMOF-1 at 80 K,
    (b) IRMOF-1 at 102 K,
    (c) IRMOF-1 at 130 K,
    (d) IRMOF-8 at 80 K,
    (e) IRMOF-8 at 102 K,
    (f) IRMOF-8 at 130 K. Adsorption isotherms at 92 K can be found in \textbf{Figure \ref{fig:ModParam_Isotherms_IRMOF-1,8}}.}
    \label{fig:Isotherms_IRMOFs-1,8}
\end{figure}

Increasing CO$_2$ loading produces two primary consequences: (i) a reduction in the maximum CH$_4$ capacity through competition for pore volume and adsorption sites, and (ii) a strengthening of effective adsorption interactions that decreases the jump pressure, $P_j$, and simultaneously increases the Henry's constant, $K_{\rm H}$. By strengthening these interactions, the first layer is adsorbed at lower pressures. Adsorption of the second layer is thermodynamically less stable than the filled pore due to strengthened CH$_4$-CH$_4$ interactions. In addition, the distribution of CO$_2$ may introduce wall effects that alter the adsorption of CH$_4$.

The decrease in the saturation CH$_4$ uptake with increasing CO$_2$ loading can be attributed somewhat trivially to the occupation of the pore by the latter, given its stronger enthalpy of adsorption \cite{GomezGualdron2014, Diaz2026}: $\Delta N_{sat}$ decreases approximately linearly with $N_{\rm CO_2}$, with a slope  $\sim$ 0.85 in IRMOF-1 and 0.89 in IRMOF-8, whereas the Bondi van der Waals radii ratio is $V_{\rm CH_4}/V_{\rm CO_2} \approx 0.85$ \cite{Bondi1964,Zhao2003}. The competition for adsorption sites can be observed in more detail in the density maps of CH$_4$ and CO$_2$ (see \textbf{SI-2} in the Supporting Information).

\begin{figure}[t]
    \centering
    % First subfigure
    \begin{subfigure}[b]{0.45\textwidth}
        \begin{overpic}[width=\textwidth]{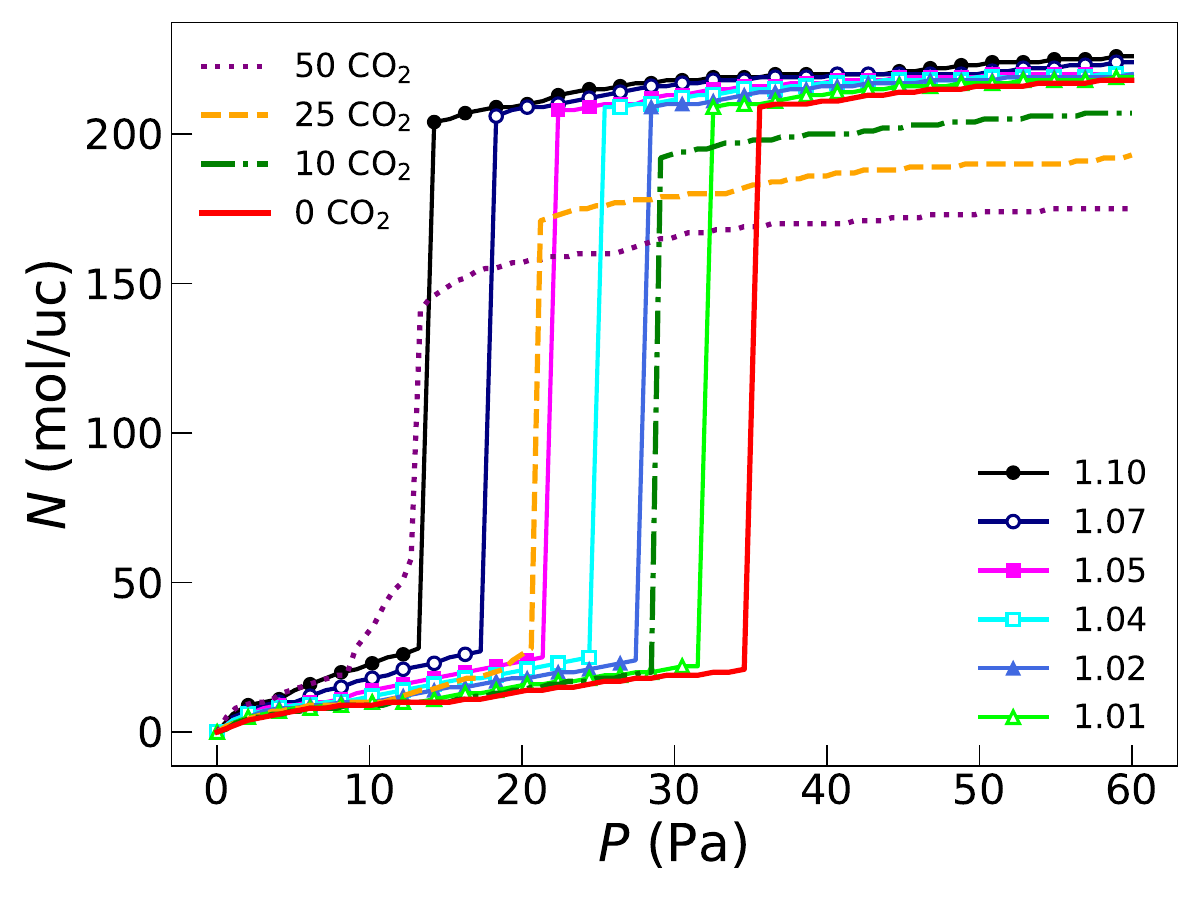}
            \put(0,70){\textbf{a)}} % 2% from left, 95% from bottom
        \end{overpic}
        \label{fig:Isotherm_Comparison_IRMOF-1}
    \end{subfigure}
    \hfill
    % Second subfigure
    \begin{subfigure}[b]{0.45\textwidth}
        \begin{overpic}[width=\textwidth]{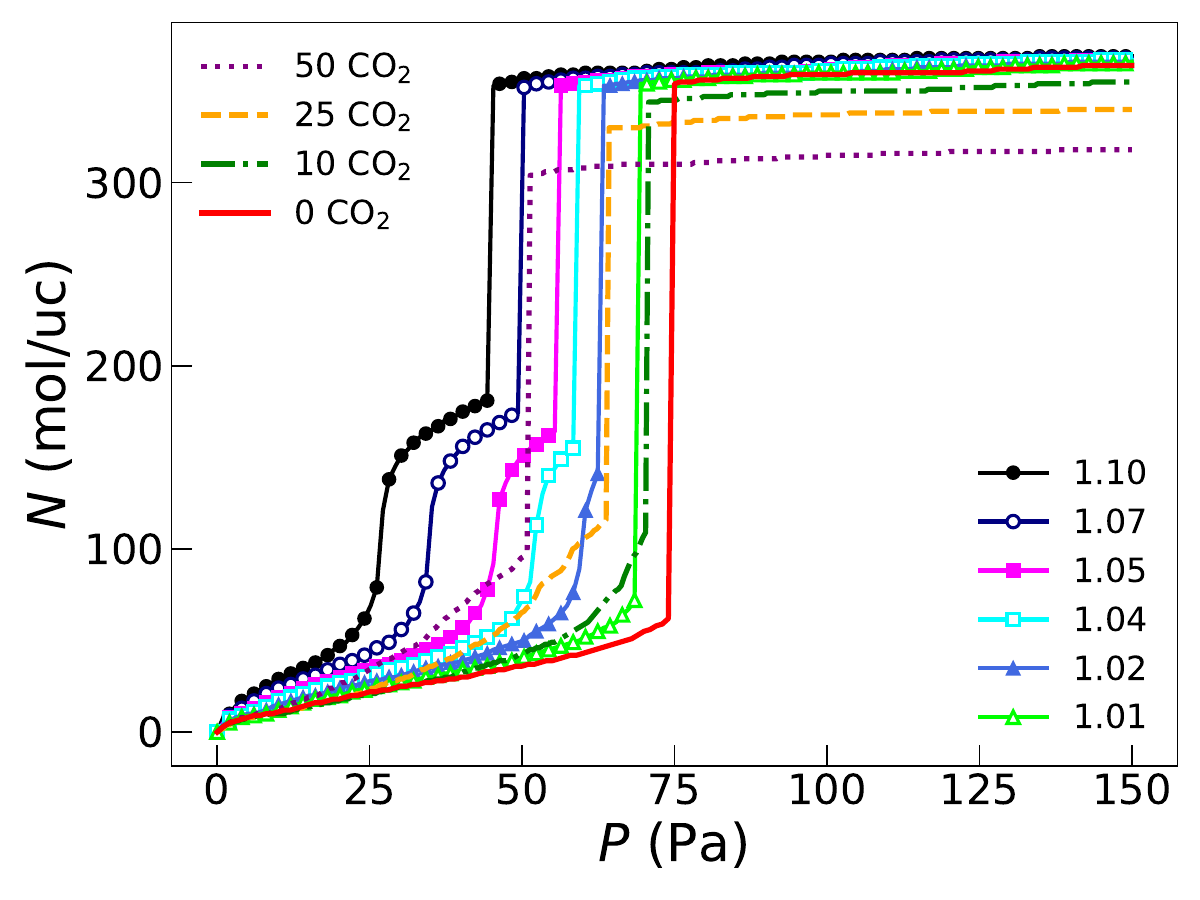}
            \put(0,70){\textbf{b)}}
        \end{overpic}
        \label{fig:Isotherm_Comparison_IRMOF-8}
    \end{subfigure}
    \caption{
    (a) Adsorption isotherm comparison for CH$_4$ between systems with added CO$_2$ and modified interactions ($\lambda$) in: (a) IRMOF-1 and (b) IRMOF-8.
    }
    \label{fig:ModParam_Isotherms_IRMOF-1,8}
\end{figure}
 
The reduction of the jump pressure, $P_j$, is associated with the presence of CO$_2$, which appears to effectively strengthen the adsorption potential experienced by CH$_4$. To test this hypothesis we performed additional simulations of pure CH$_4$, but where the interaction strength between CH$_4$ and the framework atoms was artificially increased by a scaling factor, $\lambda \in [1.00,1.10$]. \textbf{Figure \ref{fig:ModParam_Isotherms_IRMOF-1,8}} shows the isotherms obtained with modified parameters alongside those for systems containing 0, 10, 25, and 50 CO$_2$ molecules in both IRMOF-1 and IRMOF-8 at 92 K. A simple scaling argument based on the Monte Carlo transition probabilities (Eqs. \ref{eq:p_insertion}, \ref{eq:p_deletion}) \cite{Fitzgerald1999, Errington2003, Shen2004, Siderius2013} suggests that the chemical potential at the transition point depends approximately linearly on the energy of an adsorbed molecule. Thus, the jump pressure is expected to result in a qualitative exponential dependence $P_j \sim e^{-\beta \lambda U}$ in the case of the systems with modified interaction parameters \cite{Khalfaoui2003, BenTorkia2014}.

\begin{figure}[t]
    \centering
\includegraphics[width=1\textwidth]{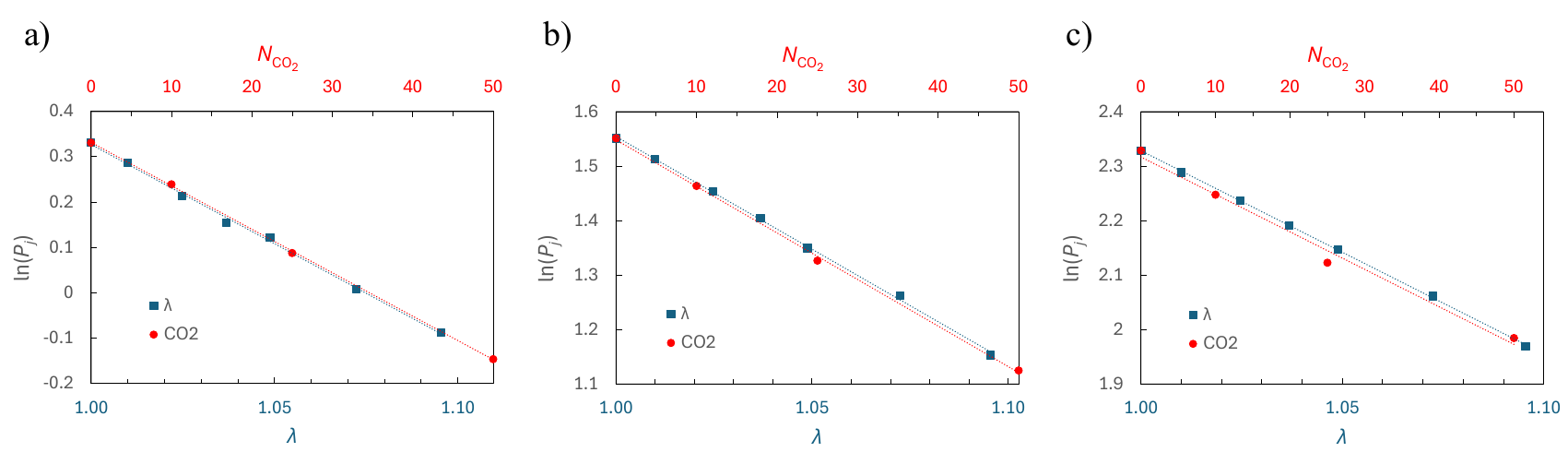}
\caption{
    Relation between jump pressure $P_j$ and the interaction strength parameter $\lambda$ (bottom axis) or the number of added CO$_2$ molecules (top axis) at 
    (a) 80 K,
    (b) 92 K, and
    (c) 102 K.}
    \label{fig:lnpj2_IRMOF-1}
\end{figure}

\textbf{Figure \ref{fig:lnpj2_IRMOF-1}} shows this behavior clearly, with a remarkably similar dependence on the number of CO$_2$ molecules $N_{\rm CO_2}$, confirming the hypothesis above.  In fact, comparison of slopes in \textbf{Figure \ref{fig:lnpj2_IRMOF-1}} yields that each added CO$_2$ molecule correspond to a $\sim 0.2\%$ increase in the interaction strength. An Arrhenius-type analysis ($\ln(P_j)$ vs. $1/T$, see \textbf{SI-3}) results in $U_0 \approx 14$ kJ/mol with a correlation factor $R^2 = 1$, as expected from typical enthalpies of adsorption for CH$_4$ in IRMOF materials (10.5-14.5 kJ/mol) \cite{GomezGualdron2014}. It is quite remarkable that the addition of a few CO$_2$ molecules drastically changes the adsorption behavior, possibly resulting in a significant change in the number of adsorbed CH$_4$ molecules by lowering the transition pressure $P_j$ ({\textbf{Figures \ref{fig:Isotherms_IRMOFs-1,8}, \ref{fig:ModParam_Isotherms_IRMOF-1,8}}}).

%%%%%%%%%%%%%%%%%%%%%  
\begin{figure}[ht]
    \centering
    \begin{subfigure}[b]{0.6\textwidth}
        \begin{overpic}[width=\textwidth]{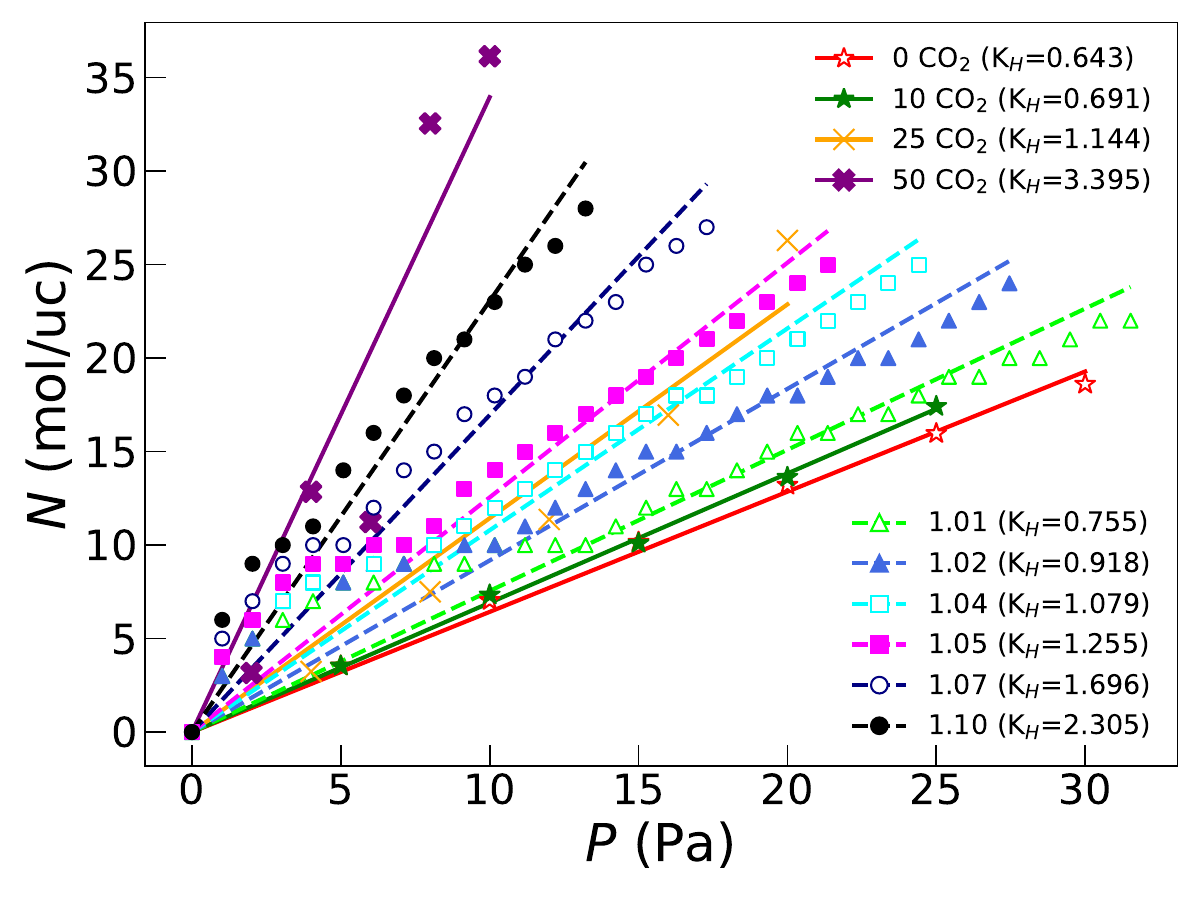}
        \end{overpic}
        \label{fig:Henry's_Constant_92K}
    \end{subfigure}
    \caption{
    Low pressure region of the adsorption isotherms for CH$_4$ in IRMOF-1 at 92 K. Linear fits result in effective Henry's law constants $K_{\rm H}$ in mol/uc\,Pa. Solid lines: systems with added CO$_2$; dashed lines: modified interaction ($\lambda$). 
    }
    \label{fig:kH}
\end{figure}

\textbf{Figure \ref{fig:kH}} shows the low pressure regions of the CH$_4$ adsorption isotherms in IRMOF-1 at 92 K for systems containing $N_{\rm CO_2} = $ 0, 10, 25, and 50  alongside those obtained from simulations with artificially modified interaction parameters $\lambda \in [1.00,1.10]$. Grand Canonical Monte Carlo (GCMC) simulations were employed in this region to improve statistical sampling at low CH$_4$ loadings. Henry's law constants $K_{\rm H} = \partial N/\partial P$ for each curve are calculated from best linear fits and shown in the figure. In the low-density regime, $K_{\rm H}$ is expected to depend exponentially on the adsorption potential \cite{Burress2009}.  Consequently, a linear relationship is observed between $\ln (K_{\rm H})$ and the interaction-strength scaling parameter, $\lambda$ (see \textbf{SI-3}). 
As in the case for the jump pressure, we find a qualitatively similar dependence of $K_{\rm H}$ with the number of CO$_2$ molecules added into the system, e.g.,  $N_{\rm CO_2} = 25$ behaves similarly to $\lambda \sim 1.044$.  

The agreement between the dependence of $P_j$ and $K_{\rm H}$ on both the interaction strength (set by $\lambda$) and the addition of CO$_2$ molecules support the idea that the structural transformation is controlled primarily by the energetics of the initial adsorption layer, rather than by packing effects that emerge at higher loadings. 

%%%%%%%%%%%%%%%%%%%%%%%%%%%%%%%%%%%%%%%%%%%%%%%%%%%

\begin{figure}[h!]
    \centering
    % Row 1
    \begin{subfigure}[b]{0.48\textwidth}
        \begin{overpic}[width=\textwidth]{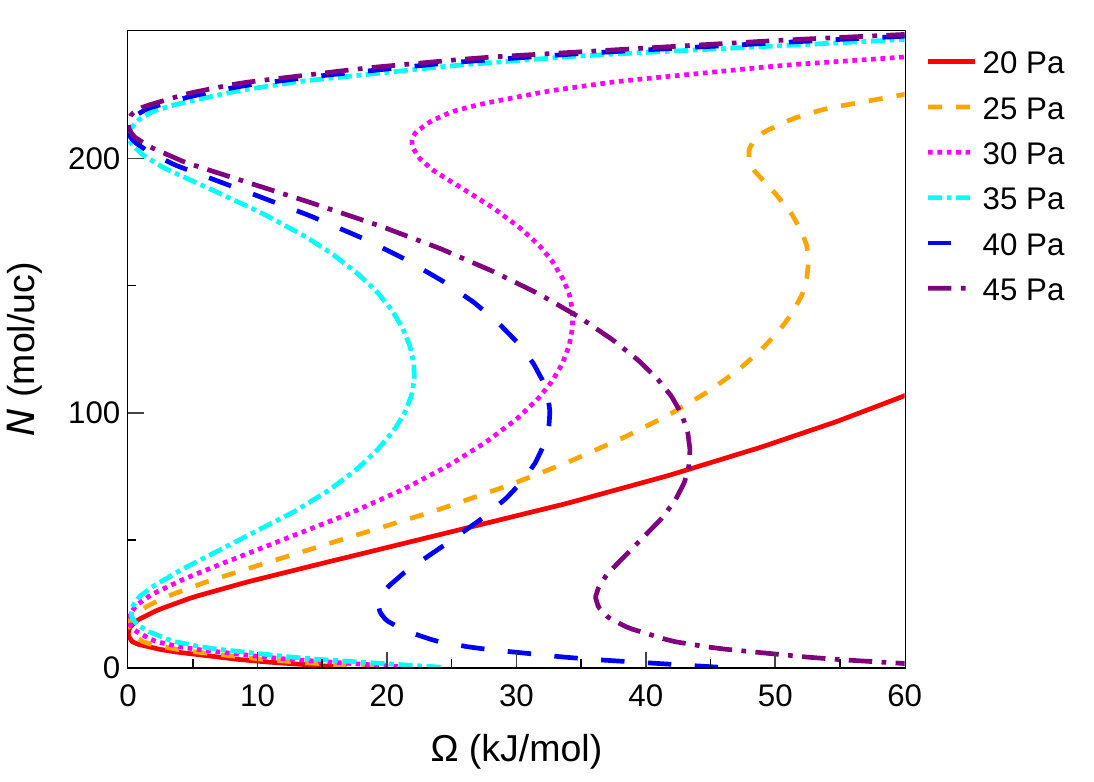}
            \put(0,70){\textbf{a)}}
        \end{overpic}
    \end{subfigure}
    \hfill
    \begin{subfigure}[b]{0.48\textwidth}
        \begin{overpic}[width=\textwidth]{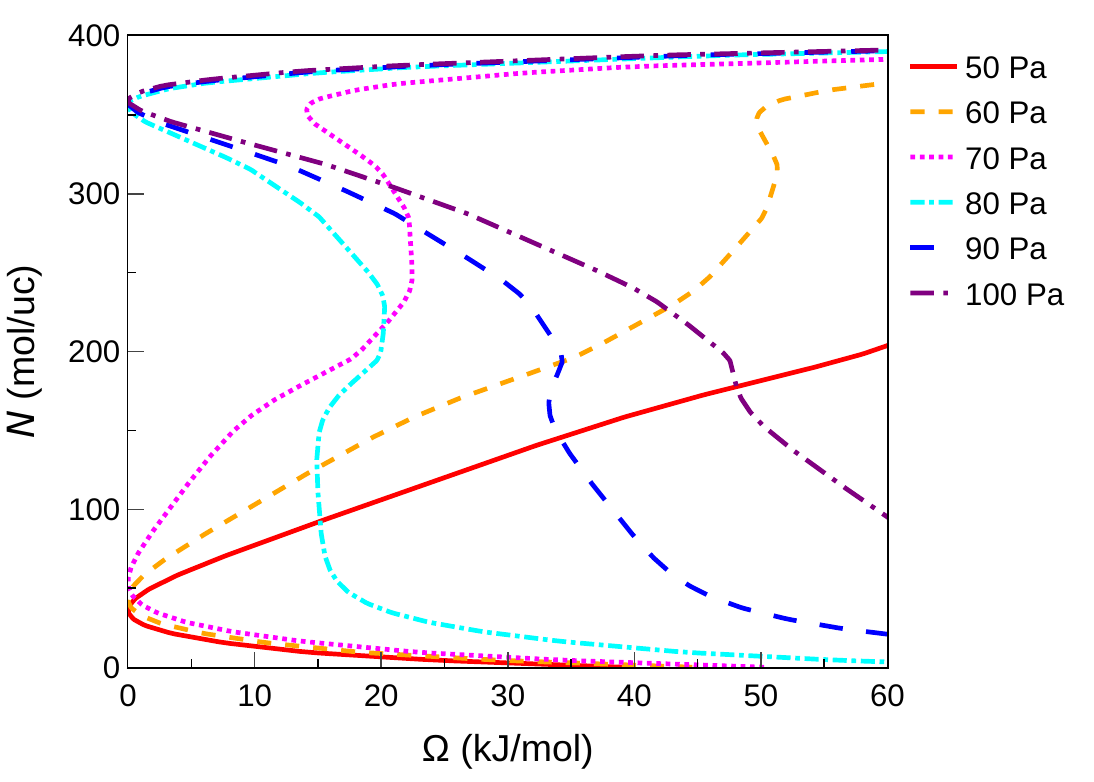}
            \put(0,70){\textbf{b)}}
        \end{overpic}
    \end{subfigure}
    \vspace{0.5cm}
    % Row 2
    \begin{subfigure}[b]{0.48\textwidth}
        \begin{overpic}[width=\textwidth]{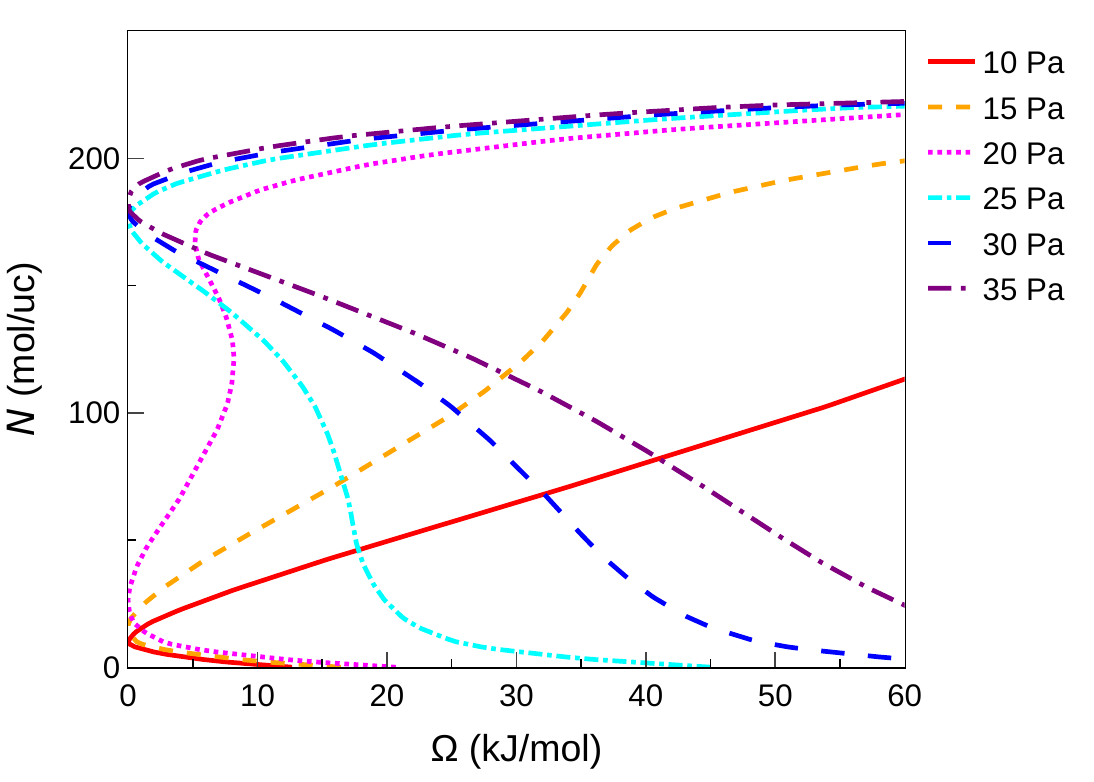}
            \put(0,70){\textbf{c)}}
        \end{overpic}
    \end{subfigure}
    \hfill
    \begin{subfigure}[b]{0.48\textwidth}
        \begin{overpic}[width=\textwidth]{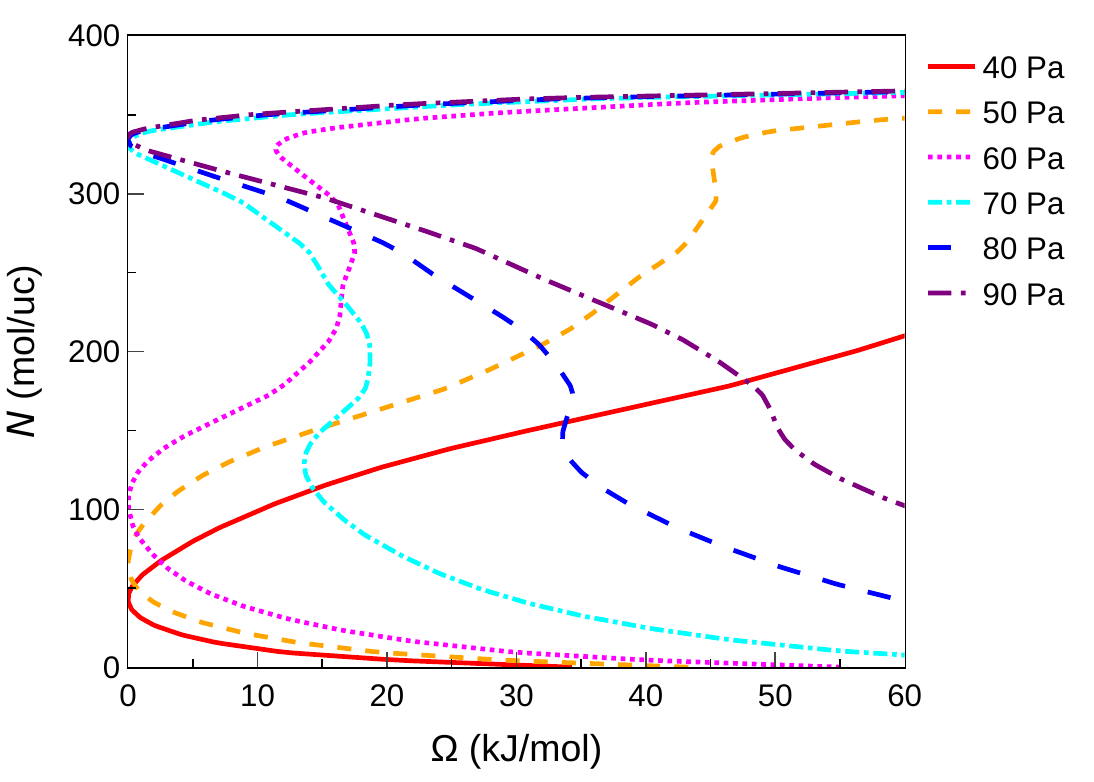}
            \put(0,70){\textbf{d)}}
        \end{overpic}
    \end{subfigure}
    \vspace{0.5cm}
    % Row 3
    \begin{subfigure}[b]{0.48\textwidth}
        \begin{overpic}[width=\textwidth]{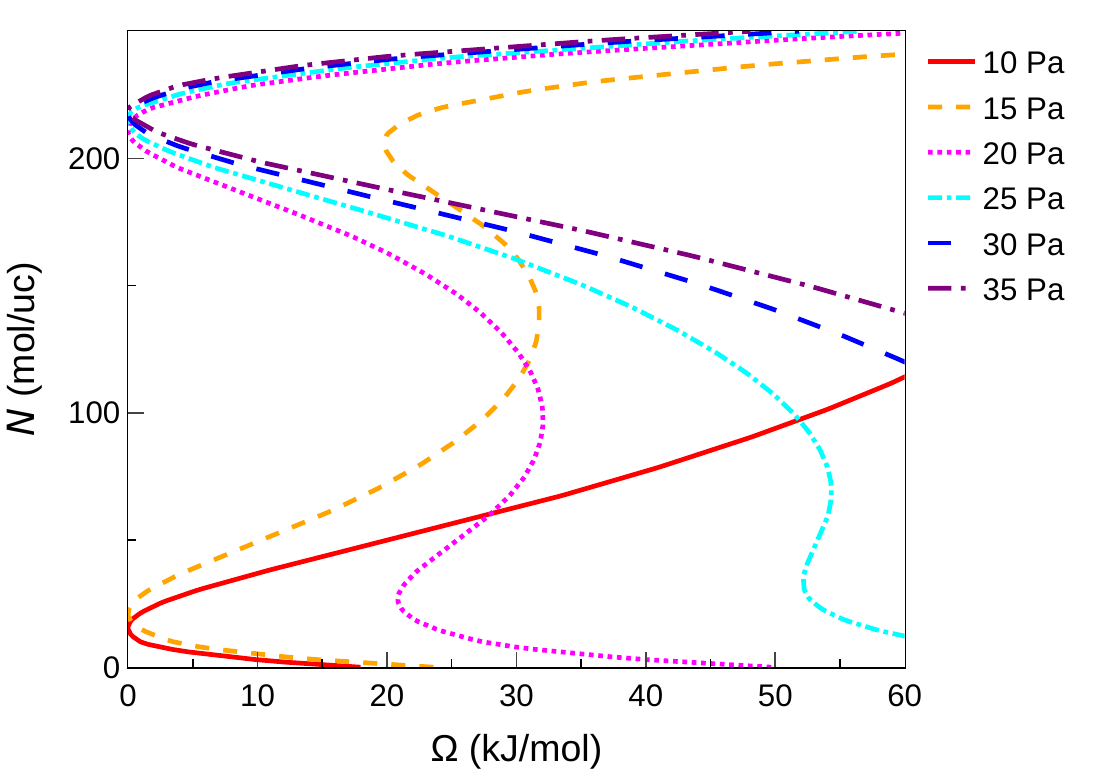}
            \put(0,70){\textbf{e)}}
        \end{overpic}
    \end{subfigure}
    \hfill
    \begin{subfigure}[b]{0.48\textwidth}
        \begin{overpic}[width=\textwidth]{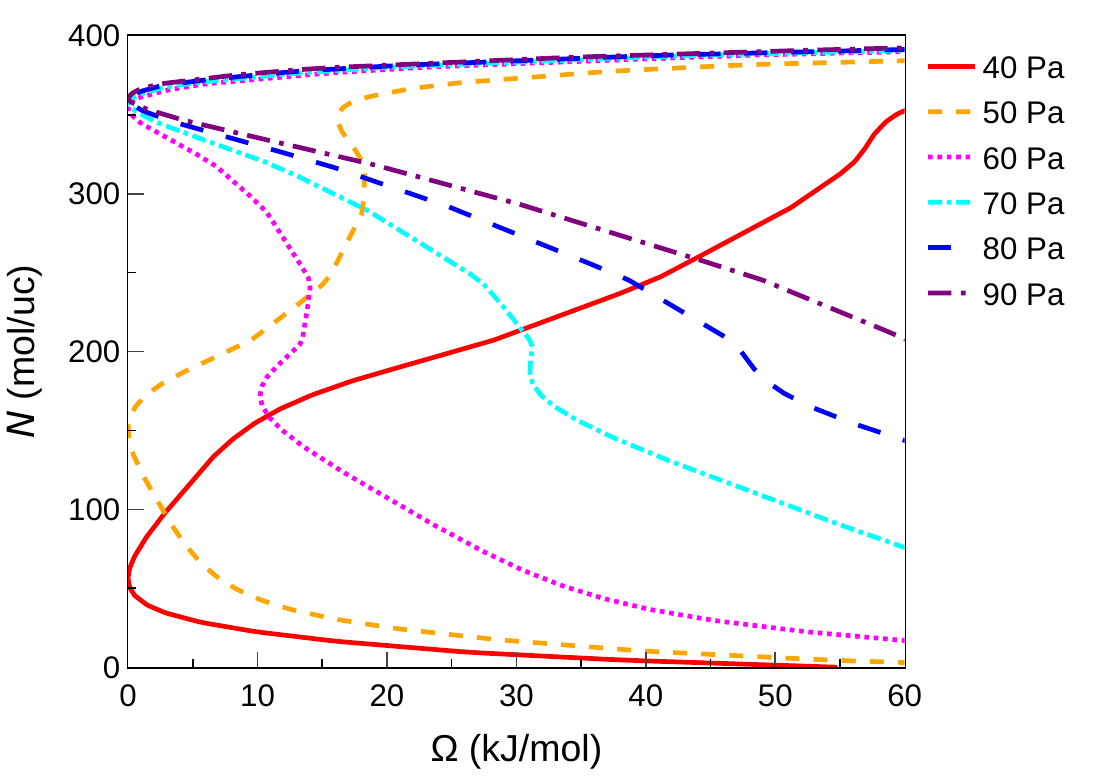}
            \put(0,70){\textbf{f)}}
        \end{overpic}
    \end{subfigure}
    \caption{Free energy profiles for CH$_4$ in IRMOF-1 (left column) and IRMOF-8 (right column) at 92 K for different number of CO$_2$ molecules ($N_{\rm CO_2}$) and interaction strengths ($\lambda$). (a) and (b) $N_{\rm CO_2} = 0$, $\lambda = 1.00$. (c) and (d) $N_{\rm CO_2} = 25$, $\lambda = 1.00$. (e) and (f) $N_{\rm CO_2} = 0$, $\lambda = 1.05$. Figure \textbf{SI-4} contains plots of the free energy plotted as a function of pressure and uptake.}
    \label{fig:Free_Energies_92K}
\end{figure}

%%%%%%%%%%%%%%%%%%%%%%%%%%%%% 

\textbf{Figure \ref{fig:Free_Energies_92K}} compares the free energy profiles for CH$_4$ adsorption at 92 K in IRMOF-1 and IRMOF-8. The top row shows the reference systems containing only CH$_4$. The middle row illustrates the effect of introducing 25 CO$_2$ molecules while leaving the interaction parameters unchanged. The bottom row shows the corresponding pure CH$_4$ systems in which the CH$_4$-MOF LJ interaction strength was artificially increased by a multiplicative factor $\lambda = 1.05$. Together, these profiles allow the effects of CO$_2$ addition and interaction-strength modification to be compared directly. 

We observe that both the addition of CO$_2$ molecules ($N_{\rm CO_2}$) and the strengthening of the CH$_4$-MOF interactions ($\lambda$) modify the underlying free energy landscape lowering the pressure $P_j$ at which the global minimum of the free energy (the equilibrium state) shifts from a low-occupation to a high-occupation state, as observed earlier in \textbf{Figures \ref{fig:Isotherms_IRMOFs-1,8}, \ref{fig:ModParam_Isotherms_IRMOF-1,8}, \ref{fig:lnpj2_IRMOF-1}}. Additionally, in both cases there is a reduction in the height of the free energy barriers separating the local minima. 

In the case of IRMOF-1, where only low- and high-occupation states are present, this reduction in barrier height facilitates a shift of the transition to lower pressures, which is reflected in an increased Henry's constant $K_{\rm H}$ (\textbf{Figure \ref{fig:kH}}). In contrast, IRMOF-8 exhibits a more complex landscape with multiple metastable states separated by two free energy barriers \cite{Dorhauer2025}. The reduction of the energy barriers effectively eliminates the intermediate metastable state, and causes the two low-occupation minima to merge, resulting in the  intermediate region in the adsorption isotherms (\textbf{Figure \ref{fig:ModParam_Isotherms_IRMOF-1,8}}).

Looking back at the point where the occupation jumps in \textbf{Figures \ref{fig:Isotherms_IRMOFs-1,8}, \ref{fig:ModParam_Isotherms_IRMOF-1,8}} the transition appears to qualitatively occur at a similar CH$_4$ occupation. This indicates that the structural transition observed in the CH$_4$ system happens at a particular point where the CH$_4$-CH$_4$ interactions cooperatively lead to a reconfiguration of the system as previously observed \cite{Kuchta2017, Mazur2022, Dorhauer2025, Stankiewicz2026}. As Henry's constant increases with $N_{\rm CO_2}$ and $\lambda$ (\textbf{Figure \ref{fig:kH}}), $P_j$ is reduced.

For IRMOF-8 the scenario is more complex: the intermediate metastable state alters the manner in which the modified parameter isotherms (\textbf{Figure \ref{fig:ModParam_Isotherms_IRMOF-1,8}b}) evolve. The difference can be attributed to the fact that CO$_2$ doesn't just displace CH$_4$ molecules, but substantially alters the corresponding free energy landscapes governing CH$_4$ adsorption. While the underlying mechanism is analogous to that in IRMOF-1, the presence of an additional barrier results in a two-step transition, characterized by two smaller, less abrupt jumps in the isotherm rather than a single sharp transition. Importantly, this behavior is not replicated simply by increasing the CO$_2$ concentration. In addition to strengthening CH$_4$–MOF interactions, CO$_2$ molecules occupy pore volume and compete for adsorption sites, thereby reducing the maximum CH$_4$ uptake. As a result, the medium-density metastable state is not merely destabilized but is effectively eliminated due to the reduced saturation capacity. The disappearance of the intermediate minimum in IRMOF-8 provides direct evidence that the observed changes in the adsorption isotherm originate from a restructuring of the underlying free energy landscape rather than from a simple shift in adsorption affinity \cite{Dorhauer2025}.

The results demonstrate that added CO$_2$ molecules introduce additional attractive Lennard-Jones interactions that deepen the effective adsorption potential experienced by CH$_4$, with each CO$_2$ molecule (per unit cell) contributing approximately a 0.2\% increase in effective potential depth. Thus, controlled CO$_2$ addition provides a means of tuning CH$_4$ adsorption strength while simultaneously modifying available pore volume and the underlying free energy landscape. This approach could be particularly relevant to natural gas storage, where adsorption strength must balance high uptake with a sufficiently favorable chemical potential for subsequent desorption. More broadly, controlled co-adsorption offers a potential route for tuning adsorption-induced transformations in nanoporous materials without modifying the framework itself.

\section{Summary and Conclusions}

In this work, we investigated the influence of CO$_2$ impurities on the structural transformation of CH$_4$ in IRMOF-1 and IRMOF-8 using Transition-Matrix Monte Carlo simulations. Increasing the CO$_2$ concentration produced two systematic effects: a reduction in CH$_4$ saturation uptake, and a strengthening of the effective adsorption potential, leading to an increased Henry's constant, and subsequently a shift of the structural transformation to lower pressures. The decrease in maximum CH$_4$ uptake is attributed primarily to competition with CO$_2$ for available pore volume and its quantitative decrease is consistent with their comparable effective molecular volumes. In the case of IRMOF-8 this competition causes the low- and medium-density states to merge, ultimately eliminating the intermediate metastable state \cite{Dorhauer2025}.

The observed reduction in transition pressure is closely associated with an increase in the Henry's constant, indicating stronger effective interactions within the adsorbed phase. Analysis of the free energy landscapes reveals that CO$_2$ lowers the barriers separating low-, medium-, and high-density states, allowing the critical uptake required for the structural transformation to be reached at lower pressures. Additional simulations employing artificially strengthened CH$_4$-MOF interactions reproduced the same qualitative trends, supporting the interpretation that CO$_2$ acts not only as a competing adsorbate but also as an effective modifier of the interactions governing CH$_4$ adsorption and phase behavior.

The exponential dependence of the transition pressure on CO$_2$ loading further implies that the reduction in $P_j$ is substantially greater than the corresponding loss in CH$_4$ storage capacity, which decreases linearly with increasing CO$_2$ concentration. There are regions of pressure where CH$_4$ is normally at a low-density state and the addition of a few CO$_2$ molecules directly leads to the system transitioning to a high-density state, thus a significant increase in the CH$_4$ uptake. 

More broadly, these results demonstrate that relatively small concentrations of impurities can substantially alter the free energy landscape of confined fluids and, consequently, the thermodynamics of adsorption-induced phase transformations. These findings suggest a potential strategy for tuning adsorption behavior through controlled modification of adsorbate composition rather than framework chemistry alone. While the present study focuses on subcritical CH$_4$/CO$_2$ mixtures in IRMOF-1 and IRMOF-8, the underlying mechanism is expected to extend to other nanoporous systems exhibiting cooperative adsorption phenomena. Combining controlled co-adsorption with framework modifications designed to reduce hysteresis and improve loading capacity \cite{Prelesnik2026} may provide complementary strategies for optimizing gas storage and separation. However, further investigation is needed to assess the practical benefits and limitations of these approaches with respect to applications to gas storage and separation.

\begin{center}
{\bf Supplemental Information and Input Files available upon request.}
\end{center}

\clearpage

%%%%%%%%%%%%%%%%%%%%%%%%%%%%%%%%%%%%%%%%%%%%%%%%%%%%%%%%%%%%%%%%%%%%%%%%%%%%%%%
\bibliographystyle{unsrtnat}
\bibliography{references}

%%%%%%%%%%%%%%%%%%%%%%%%%%%%%%%%%%%%%%%%%%%%%%%%%%%%%%%%%%%%%%%%%%%%%%%%%%%%%%%
\end{document}